\documentclass[12pt]{article}
\usepackage{amsfonts}
\usepackage{amssymb}

\def\hybrid{\topmargin -20pt    \oddsidemargin 0pt
        \headheight 0pt \headsep 0pt
        \textwidth 6.35in       
        \textheight 9.15in       
        \marginparwidth .875in
        \parskip 5pt plus 1pt   \jot = 1.5ex}

\hybrid

\def\baselinestretch{1.2}

\catcode`\@=11

\def\marginnote#1{}
\newcount\hour
\newcount\minute
\newtoks\amorpm
\hour=\time\divide\hour by60
\minute=\time{\multiply\hour by60 \global\advance\minute by-\hour}
\edef\standardtime{{\ifnum\hour<12 \global\amorpm={am}%
        \else\global\amorpm={pm}\advance\hour by-12 \fi
        \ifnum\hour=0 \hour=12 \fi
        \number\hour:\ifnum\minute<10 0\fi\number\minute\the\amorpm}}
\edef\militarytime{\number\hour:\ifnum\minute<10 0\fi\number\minute}

\def\draftlabel#1{{\@bsphack\if@filesw {\let\thepage\relax
   \xdef\@gtempa{\write\@auxout{\string
      \newlabel{#1}{{\@currentlabel}{\thepage}}}}}\@gtempa
   \if@nobreak \ifvmode\nobreak\fi\fi\fi\@esphack}
        \gdef\@eqnlabel{#1}}
\def\@eqnlabel{}
\def\@vacuum{}
\def\draftmarginnote#1{\marginpar{\raggedright\scriptsize\tt#1}}

\def\draft{\oddsidemargin -.5truein
        \def\@oddfoot{\sl preliminary draft \hfil
        \rm\thepage\hfil\sl\today\quad\militarytime}
        \let\@evenfoot\@oddfoot \overfullrule 3pt
        \let\label=\draftlabel
        \let\marginnote=\draftmarginnote
   \def\@eqnnum{(\theequation)\rlap{\kern\marginparsep\tt\@eqnlabel}%
\global\let\@eqnlabel\@vacuum}  }

\def\preprint{\twocolumn\sloppy\flushbottom\parindent 2em
        \leftmargini 2em\leftmarginv .5em\leftmarginvi .5em
        \oddsidemargin -.5in    \evensidemargin -.5in
        \columnsep .4in \footheight 0pt
        \textwidth 10.in        \topmargin  -.4in
        \headheight 12pt \topskip .4in
        \textheight 6.9in \footskip 0pt
        \def\@oddhead{\thepage\hfil\addtocounter{page}{1}\thepage}
        \let\@evenhead\@oddhead \def\@oddfoot{} \def\@evenfoot{} }

\def\numberbysection{\@addtoreset{equation}{section}
        \def\theequation{\thesection.\arabic{equation}}}

\def\underline#1{\relax\ifmmode\@@underline#1\else
        $\@@underline{\hbox{#1}}$\relax\fi}

\def\titlepage{\@restonecolfalse\if@twocolumn\@restonecoltrue\onecolumn
     \else \newpage \fi \thispagestyle{empty}\c@page\z@
        \def\thefootnote{\fnsymbol{footnote}} }

\def\endtitlepage{\if@restonecol\twocolumn \else \newpage \fi
        \def\thefootnote{\arabic{footnote}}
        \setcounter{footnote}{0}}  

\catcode`@=12
\relax

\def\figcap{\section*{Figure Captions\markboth
        {FIGURECAPTIONS}{FIGURECAPTIONS}}\list
        {Figure \arabic{enumi}:\hfill}{\settowidth\labelwidth{Figure
999:}
        \leftmargin\labelwidth
        \advance\leftmargin\labelsep\usecounter{enumi}}}
 \relax
\def\tablecap{\section*{Table Captions\markboth
        {TABLECAPTIONS}{TABLECAPTIONS}}\list
        {Table \arabic{enumi}:\hfill}{\settowidth\labelwidth{Table
999:}
        \leftmargin\labelwidth
        \advance\leftmargin\labelsep\usecounter{enumi}}}
 \relax
\def\reflist{\section*{References\markboth
        {REFLIST}{REFLIST}}\list
        {[\arabic{enumi}]\hfill}{\settowidth\labelwidth{[999]}
        \leftmargin\labelwidth
        \advance\leftmargin\labelsep\usecounter{enumi}}}
 \relax
\makeatletter
\newcounter{pubctr}
\def\publist{\@ifnextchar[{\@publist}{\@@publist}}
\def\@publist[#1]{\list
        {[\arabic{pubctr}]\hfill}{\settowidth\labelwidth{[999]}
        \leftmargin\labelwidth
        \advance\leftmargin\labelsep
        \@nmbrlisttrue\def\@listctr{pubctr}
        \setcounter{pubctr}{#1}\addtocounter{pubctr}{-1}}}
\def\@@publist{\list
        {[\arabic{pubctr}]\hfill}{\settowidth\labelwidth{[999]}
        \leftmargin\labelwidth
        \advance\leftmargin\labelsep
        \@nmbrlisttrue\def\@listctr{pubctr}}}
 \relax
\makeatother
\newskip\humongous \humongous=0pt plus 1000pt minus 1000pt

\newif\ifdtup

\relax

\def\be{\begin{equation}}
\def\ee{\end{equation}}
\def\ba{\begin{eqnarray}}
\def\ea{\end{eqnarray}}

\def\del{\partial}

\def\r{\rho}
\def\a{\alpha}

\def\b{\beta}

\def\g{\gamma}
\def\G{\Gamma}
\def\d{\delta}

\def\e{\epsilon}

\def\th{\theta}

\def\m{\mu}
\def\n{\nu}
\def\om{\omega}
\def\Om{\Omega}
\def\l{\lambda}

\def\s{\sigma}

\def\bs{\bigskip}
\def\no{\noindent}

\def\qq{\qquad}

\def\IR{\relax{\rm I\kern-.18em R}}
\def\II{\relax{\rm 1\kern-.35em1}}
\renewcommand{\theequation}{\thesection.\arabic{equation}}
\csname @addtoreset\endcsname{equation}{section}

\def \ha {{1\over 2}}

\def \ov {\over}

\def\IR{\relax{\rm I\kern-.18em R}}
\def\inv{^{\raise.15ex\hbox{${\scriptscriptstyle -}$}\kern-.05em 1}}

\begin{document}


\newcommand{\beq}{\begin{equation}}
\newcommand{\eeq}[1]{\label{#1}\end{equation}}
\newcommand{\ber}{\begin{eqnarray}}
\newcommand{\eer}[1]{\label{#1}\end{eqnarray}}
\newcommand{\eqn}[1]{(\ref{#1})}
\begin{titlepage}
\begin{center}

\hfill NEIP-03-004\\
\vskip -.1 cm
\hfill hep--th/0309147\\

\vskip .5in

{\Large \bf  Supersymmetric quantum mechanics \\ from wrapped branes}

\vskip 0.4in

{\bf Rafael Hern\'andez$^1$}\phantom{x} and\phantom{x}
 {\bf Konstadinos Sfetsos}$^2$ 
\vskip 0.1in

${}^1\!$
Institut de Physique, Universit\'e de Neuch\^atel\\
Breguet 1, CH-2000 Neuch\^atel, Switzerland\\
{\footnotesize{\tt rafael.hernandez@unine.ch}}

\vskip .2in

${}^2\!$
Department of Engineering Sciences, University of Patras\\
26110 Patras, Greece\\
{\footnotesize{\tt sfetsos@des.upatras.gr}}\\

\end{center}

\vskip .3in

\centerline{\bf Abstract}

\no
We explicitly construct a solution of eight-dimensional gauged supergravity 
representing D6-branes wrapped 
on six-cycles inside Calabi--Yau fourfolds. 
The solution preserves two supercharges
and asymptotically is a cone with the coset space $SU(2)^4/U(1)^3$ as its base.
It is shown to correspond to an M-theory compactification on a Calabi--Yau 
manifold with $SU(5)$ holonomy and we discuss in detail its geometrical 
and topological features.
We also construct a family of related higher dimensional metrics 
having $SU(n\!+\!1)$ holonomy, which of course have no brane interpretation.

\noindent

\vskip .4in
\noindent

\end{titlepage}
\vfill
\eject

\def\baselinestretch{1.2}


\baselineskip 20pt

\section{Introduction}

D6-branes have a purely geometrical origin in eleven dimensions as the 
Kaluza--Klein monopole. When 
the amount of supersymmetry on their worldvolume is reduced by wrapping them on supersymmetric cycles 
they admit an eleven-dimensional description in terms of compactifications of M-theory on manifolds with reduced holonomy 
\cite{Gomis}. Keeping some unbroken supersymmetry as the brane wraps a cycle requires coupling the theory 
to an external R-current, which is then related to the spin 
connection on the cycle. The 
resulting theory is a topologically twisted field theory \cite{BSV}. 
At a more technical level, gauged supergravities 
have provided the adequate arena to perform the twist relating the gauge and 
spin connections \cite{MN} (see for instance \cite{Gauntlett} for a review). 
  
In this way compactifications of M-theory on manifolds with reduced holonomy arise 
as the local eleven-dimensional version of backgrounds in eight-dimensional gauged supergravity 
describing D6-branes wrapped on diverse supersymmetric cycles. M-theory on Calabi--Yau threefolds 
corresponds to D6-branes wrapped on supersymmetric two-cycles inside twofolds \cite{EN,EPR3}, 
compactification on Calabi--Yau fourfolds comes from D6-branes wrapped on four-cycles inside 
threefolds \cite{GM}, compactification on manifolds with $G_2$ holonomy arises as the 
eleven-dimensional description of D6-branes wrapped on supersymmetric three-cycles 
inside Calabi--Yau threefolds \cite{EN,HS,EPR2}, and 
M-theory on eight-manifolds with Spin(7) holonomy corresponds to D6-branes wrapped on 
four-cycles inside manifolds with $G_2$ holonomy \cite{Hernandez}. The deformation of this 
purely geometrical solutions by non-trivial background fluxes has also been extensively 
studied \cite{Gursoy}-\cite{Barcelona2} (a more complete list of references on branes of different 
dimension wrapped on various supersymmetric cycles can 
be found in \cite{HSleuven}). 
  
In this letter we will consider the case of D6-branes wrapped on six-cycles
inside Calabi--Yau fourfolds. The 
remaining theory on the worldvolume of the branes will be a supersymmetric quantum mechanics 
with two supercharges, and when lifted to eleven dimensions the solution will 
correspond to M-theory compactified on a Calabi--Yau 
manifold of $SU(5)$ holonomy. In section 2 we will present the most general 
supersymmetry preserving solution and its deformation by a non-trivial 
background flux. In section 3 we generalize our results 
to provide a family of metrics having $SU(n\!+\!1)$ holonomy 
and conclude with some remarks. 


\section{D6-branes wrapped on supersymmetric six-cycles}

Before constructing our solution let us briefly review the relevant 
sector of gauged supergravity in eight dimensions which was 
originally constructed by Salam and Sezgin \cite{Salam}
through a Scherk--Schwarz compactification of eleven-dimensional supergravity 
\cite{11sugra} on an $SU(2)$ group manifold. The field content of the theory 
consists of the metric $g_{\mu \nu}$, a dilaton 
$\Phi$, five scalars given by a unimodular 
$3 \times 3$ matrix $L_{\alpha}^{i}$ in the coset 
$SL(3, \IR)/SO(3)$ and an $SU(2)$ 
gauge potential $A_{\mu}$, all in the gravity sector, 
and a three-form and three vector fields coming from reduction of the eleven-dimensional 
three-form.\footnote{Reduction of the eleven-dimensional 
three-form also produces a scalar and three two-forms. 
However, we will set all these fields to zero.} 
In addition, on the fermion side we have 
the pseudo--Majorana spinor $\psi_{\mu}$ and the gaugino $\chi_i$.
 
The supersymmetry variations for the gaugino and the gravitino are given by 
\ba
\delta \chi_i \!\! & = \!\! &
 \frac {1}{2} (P_{\mu \; ij} + \frac {2}{3} \delta_{ij} \partial_{\mu} 
\Phi) \hat{\Gamma}^{j} 
\Gamma^{\mu} \epsilon - \frac {1}{4} e^{\Phi} F_{\mu \nu \; i} 
\Gamma^{\mu \nu} \epsilon -  \frac {g}{8} e^{-\Phi} (T_{ij} - \frac {1}{2} \delta_{ij} T) 
\epsilon^{jkl} \hat{\Gamma}_{kl} \epsilon
\nonumber \\ 
& \!\! - \!\!\!\! & \frac {1}{144} e^{\Phi} G_{\m \n \r \s} \hat{\G}_i 
\G^{\m \n \r \s} \e - \frac {1}{24} e^{-\Phi} \e^{kln} G_{\m\n n} 
(\hat \G_{kli} + 4 \hat \G_k \d_{li}) \G^{\m\n} \e  = 0 \ ,  
\nonumber \\
\delta \psi_{\l} & \!\! = \!\! & 
{\cal D}_{\l} \epsilon + \frac {1}{24} e^{\Phi} F_{\mu \nu}^{i} 
\hat{\Gamma}_i ( \Gamma_{\l}^{\: \: \mu \nu} 
- 10 \delta_{\l}^{\, \mu} \Gamma^{\nu}) \epsilon - \frac {g}{288} e^{- \Phi} \epsilon_{ijk} 
\hat{\Gamma}^{ijk} \Gamma_{\l} T \epsilon 
\nonumber \\ 
& \!\! - \!\!\!\! & \frac {1}{96} e^{\Phi} G_{\m \n \r \s} 
( \G_{\l}^{\: \: \m \n \r \s} \! - \! 4 \d^{\m}_{\: \: \l} \G^{\n \r \s} ) \e 
- \frac {1}{48} e^{-\Phi} \e^{ijk}  G_{\m \n k} \hat \G_{ij} 
( \G_{\l}^{\: \: \m \n } \! - \! 10 \d^{\m}_{\: \l} \G^{\n} ) \e= 0 \ ,
\label{susy}
\ea
where the Yang--Mills field strength is $F_{\mu \nu}^{\a}$, the covariant derivative is defined as
\begin{equation}
{\cal D}_{\l} \epsilon = \partial_{\l} \epsilon + \frac {1}{4} \omega_{\l}^{ab} \Gamma_{ab} 
\epsilon + \frac {1}{4} Q_{\l \; ij} \hat{\Gamma}^{ij} \epsilon \, ,
\end{equation}
where $P_{\gamma \, ij}$ and $Q_{\gamma \, ij}$ are, respectively, 
the symmetric and antisymmetric quantities entering 
the Cartan decomposition of the $SL(3,\mathbb{R})/SO(3)$ coset, defined through
\begin{equation}
P_{\m \, ij} + Q_{\m \, ij} \equiv L_i^{\alpha} 
( \partial_{\m} \delta_{\alpha}^{\, \beta} 
- g \, \epsilon_{\alpha \beta \delta} A^{\delta}_{\m}) L_{\beta \; j} \, ,
\label{pq}
\end{equation}
and $T_{ij}$ is the $T$-tensor defining the potential energy associated to the scalar fields,
\begin{equation}
T^{ij} \equiv L^{i}_{\alpha} L^{j}_{\beta} \delta^{\alpha \beta} \, ,
\end{equation}
with $T \equiv T_{ij} \delta^{ij}$, while $L_{\a}^{i}$ satisfy $L_{\a}^{i} L^{\a}_j = \delta^{i}_j\ ,  L_{\a}^{i} L_{\b}^{j} 
\d_{ij} = g_{\a \b}\ , L^{i}_{\a} L^{j}_{\b} g^{\a \b} = \d^{ij}$ . As usual, curved directions are 
labeled by greek indices, while flat ones are labeled by latin, and $\mu, a = 0,1, \ldots, 7$ are spacetime coordinates, 
while $\alpha, i = 8,9,10$ are in the group manifold. Note also that upper indices in the gauge field, 
$A_{\mu}^{\alpha}$, are always curved, and that the field strengths in eight dimensional curved 
space are defined as
\be
G_{\m \n \r \s} = e^{- 4 \Phi/3} e^{a}_{\: \m} e^{b}_{\: \n} e^{c}_{\: \r} e^{d}_{\: \s} F_{abcd} \ , \:\:\:\:
G_{\m \n \a\b} = e^{2 \Phi/3} e^{a}_{\: \m} e^{b}_{\: \n} 
L^i_\a L^j_\b F_{abij}=\e_{\a\b\g} G_{\m\n\g} \ ,
\label{gef}
\ee
with $G_{\m \n \a\b}$ generated by three vector fields $B_{\m\a}$ as
\be
G_{\m\n\a}= \del_\m B_{\n\a}-\del_\n B_{\m\a} \ .
\label{geff}
\ee
We should point out that the above definition ignores the contribution
to the $SU(2)$ gauge field $A^\a_\m$ which is of the form
$ \e^\g{}_{\a\b} A^\b_{\m} B_{\n\g}-(\m\leftrightarrow \n)$. 
This will be justified by the form of our ansatz which will make this term to vanish identically.
These vector fields generate also a contribution to the three three-form field strength 
$G_{\m\n\r\a}$ of the form $\e_{\a\b\g}F^\b_{\m\n} B_{\r\g}$. Again the 
absence of these terms will be justified by the form of our ansatz.
  
Let us now introduce the system under study. We will consider a D2-D6 brane system, with the 
D6-branes wrapped on supersymmetric six-cycles inside Calabi--Yau fourfolds, that is, divisors. 
As a starting point, we will take the six-cycle to be a direct product of three 
two-spheres of different radii, 
$S^2 \times \bar{S}^2 \times \tilde{S}^2$ (in an obvious notation). The 
deformation on the worl-dvolume of 
the D6-branes will then be described by a metric of the form
\be
ds_8^2 = - e^{2f} dt^2  + d\r^2 + 
\a_1^2  d\Om_2^2 + \a_2^2  d{\bar\Om}_2^2\  + \a_3^2 d{\tilde\Om}_2^2 \ ,
\label{metric}
\ee
with the line elements for the spheres (normalized to have scalar curvature equal to 2) 
\be 
d\Om^2_2 = d\th^2 +\sin^2 \th d\phi^2 \ , \qq
d{\bar\Om}^2_2 = d{\bar \th}^2 +\sin^2 {\bar \th} d{\bar \phi}^2  \ , \qq
d{\tilde\Om}^2_2 = d{\tilde \th}^2 +\sin^2 {\tilde \th} d{\tilde \phi}^2  \ ,
\label{spheres}
\ee
and $f$, $\a_1$, $\a_2$ and $\a_3$ depend only on the radial variable $\r$. 
The same dependence will also hold for all additional fields that we will turn on. 
It will be useful to introduce a triplet of Maurer--Cartan 1-forms on $S^2$,
\be
\s_1 = \sin\th d \phi\ ,\qq  \s_2 =  d\th\ , \qq \s_3=\cos\th d\phi \ ,
\ee
that obey $d\s_i=\ha \e_{ijk} \s_j\wedge \s_j$,
so that they resemble the triplet of Maurer--Cartan forms on $S^3$, although 
obviously only two of them are the independent ones. Similar
triplets, ${\bar \s}_i$ and ${\tilde \s}_i$, can also be defined on the remaining spheres, ${\bar S}^2$ 
and ${\tilde S}^2$. In the natural frame
\be
e^0 = e^f \, dt \ ,\quad e^7 = d \r \ , \quad e^i=\a_1 \s^i\ ,\quad  \bar e^i=\a_2 \bar \s^i\ ,
\quad  \tilde e^i=\a_3 \tilde \s^i\ ,\quad i=1,2 \ ,
\ee 
the expressions for the spin connection for the metric \eqn{metric} are
\ba
&& \om^{07}={df\ov d\r} e^f dt\ , \quad 
\om^{i7}={d\a_1 \ov d\r}\, \s_i\ , \quad
\bar \om^{i7}={d\a_2 \ov d\r}\, \bar \s_i\ , \quad 
\tilde \om^{i7}={d\a_1 \ov d\r}\, \tilde \s_i\, \ ,\quad i=1,2 \ ,
\nonumber\\
&& \om^{12}=\s_3\ ,\quad \om^{\bar 1\bar 2}=\bar \s_3\ ,
\quad \om^{\tilde 1\tilde 2}=\tilde \s_3\ .
\ea

The split of the six-cycle into the product in (\ref{metric}) dictates the twist, and the only 
non-vanishing component of the gauge field is
\be
A^3 =  - \frac {1}{g}(\s_3+{\bar \s}_3 + {\tilde \s}_3) \ ,
\label{gauge}
\ee
where for simplicity the overall constant has already been set to 
the value consistent with supersymmetry. The $SU(2)_R$ symmetry of the unwrapped branes is 
therefore broken to $U(1)_R$. Geometrically, the breaking of the R-symmetry happens because there 
are two normal directions to the D6-branes that are inside the Calabi--Yau fourfold; the 
R-symmetry is broken to the $U(1)_R$ on the 2-plane defined by them. 
The twist (\ref{gauge}) amounts 
to the identification of this $U(1)_R$ with a $U(1)$ subgroup in one of the $SU(2)$ factors 
in the $SO(6)$ structure group of the six-cycle. 

We will also introduce a four-form flux corresponding to D2-branes along
the two-sphere directions on the six-cycle. In order to keep democracy, we will 
turn on four-form components along all three two-spheres,
\ba
G_{x_0 \rho \s_1 \! \s_2} &=&
 Q_1 \: \frac {\a_1^2}{\a_2^2 \a_3^2} e^{-2 \Phi+f} \ ,
\nonumber\\
 G_{x_0 \rho \bar{\s}_1 \! \bar{\s}_2} &=&
Q_2 \: \frac {\a_2^2}{\a_3^2 \a_1^2}  e^{-2 \Phi+f} \ ,
\label{flux}\\
G_{x_0 \rho \tilde{\s}_1 \! \tilde{\s}_2} & = &
Q_3 \: \frac {\a_3^2}{\a_1^2 \a_2^2} e^{-2 \Phi+f} \ .
\nonumber
\ea
where all directions above are curved, 
$Q_1$, $Q_2$ and $Q_3$ are dimensionfull 
constants and the specific functional dependence is uniquely fixed by the equation of motion 
for the three-form potential. 
Among the three two-form field strengths $G_{\m\n\a}$ we will choose as only 
non-vanishing the one corresponding to $\a=3$, thus complying with the 
requirements spelled out after \eqn{geff}. Then the field strength 
solving the equation of motion for the vector $B_{\m 3}$ is 
\be
G_{x_0 \rho \a} = - Q_4 \: \frac {e^{2 \Phi+f}}{\a_1^2 \a_2^2 \a_3^2}\, \d_{\a 3}  \ ,
\label{44ff}
\ee
with $x_0$ again a curved direction and $Q_4$ dimensionfull. 
This ansatz points towards 
a particle associated with the one-form potential. 
However, this is only an artifact of the 
eight-dimensional description and in fact the flux \eqn{44ff} is 
due to D2-branes forming a bound state with the D6-branes.
We should point out that turning on just the three-form and one-form potentials
is not in general consistent with the full set of equations of motion.
Since we have set to zero the scalar and the three two-form potentials (see 
footnote 1) the corresponding equations of motion constrain the remaining 
fields.\footnote{We thank A. Paredes and A. V. Ramallo for 
a discussion on this point.} For instance, 
if we set also $G_{\m\n\a}=0$, then the non-trivial
constraints are
\be \e^{\m_1\dots \m_8} G_{\m_1\m_2\m_3\m_4} G_{\m_5\m_6\m_7\m_8}=0\ ,\qq
G_{\m\n}{}^{\r\s} F^\a_{\r\s}= 0 \ .
\label{cooo1}
\ee
In our case, using \eqn{flux} we find that \eqn{cooo1} reduces to 
$Q_1+Q_2+Q_3=0$. When we turn on $G_{\m\n\a}$ as well, the 
generalization of \eqn{cooo1} is quite complicated. However, for the case
at hand with the flux components given by \eqn{flux} and \eqn{44ff},
we obtain the simple algebraic condition 
\be
Q_1 + Q_2 + Q_3 + Q_4 = 0 \ ,
\label{zzerr}
\ee
namely, that the total flux charge vanishes.

Let's now turn to the scalars. Only the scalar corresponding 
to the unbroken $U(1)_R$ R-symmetry will survive the twisting, so 
that we will turn on one of the scalars  $L^{i}_{\a}$,
\be
L^{i}_{\a} = \hbox {diag } ( e^{\l}, e^{\l}, e^{-2 \l}) \ .
\ee
We will impose on the spinor the consistent projections 
\ba
&&\G_7 \e =  - i \G_9 \e \ , 
\nonumber \\
&&\G_1 \G_2 \e =  \bar{\G}_1 \bar{\G}_2 \e = \tilde{\G}_1 \tilde{\G}_2 \e = - \hat{\G}_1 \hat{\G}_2 \e \ .
\label{projections}
\ea
which leave in total two independent components for the spinor, so that at low energies 
we are left with a supersymmetric 
quantum mechanical model having two supercharges. 
The first projection effectively
reduces the theory along the six-cycle and the other three project into 
singlet spinors of diagonal $U(1)$'s.\footnote{A gravity dual for a quantum 
mechanics with two supercharges was also constructed in \cite{GK} using 
maximal gauged supergravity in seven dimensions to describe M-fivebranes 
wrapping a product of a three-cycle with a two-cycle.}
  
With the above ansatz and projections on the spinor, the supersymmetry 
variations for the gravitino and gaugino lead to the following equations
\ba
\frac {d \Phi}{d \r} \! & = \! & 
\frac {g}{8} e^{- \Phi} (e^{-4 \l} + 2 e^{2 \l}) - 
\frac {1}{2g} e^{\Phi -2 \l} 
\left( \frac {1}{\a_1^2} + \frac {1}{\a_2^2} + 
\frac {1}{\a_3^2} \right) 
\nonumber \\ 
& - & {(\a_1^2 Q_1 + \a_2^2 Q_2 + \a_3^2 Q_3)e^{-\Phi} - Q_4 e^{\Phi+2\l} 
\ov {2 \a_1^2\a_2^2\a_3^2}} \ , 
\nonumber \\
\frac {1}{\a_1} \frac {d \a_1}{d \r} \! & = \! & 
\frac {g}{24} e^{- \Phi} ( 2 e^{2 \l} + e^{-4 \l} ) + 
\frac {1}{6g} e^{\Phi-2\l} \left( \frac {5}{\a_1^2} - \frac {1}{\a_2^2} - 
\frac {1}{\a_3^2} \right) 
\nonumber \\ 
& - & {(-\a_1^2 Q_1 + \a_2^2 Q_2 + \a_3^2 Q_3)e^{-\Phi} + {1\ov 3}
Q_4 e^{\Phi+2\l}  \ov {2 \a_1^2\a_2^2\a_3^2}} \ , 
\nonumber\\
\frac {1}{\a_2} \frac {d \a_2}{d \r} \! & = \! & \frac {g}{24} 
e^{- \Phi} ( 2 e^{2 \l} + e^{-4 \l} ) + 
\frac {1}{6g} e^{\Phi-2\l} \left( \frac {5}{\a_2^2} - \frac {1}{\a_3^2} - 
\frac {1}{\a_1^2} \right) 
\nonumber \\ 
& - & {(\a_1^2 Q_1 - \a_2^2 Q_2 + \a_3^2 Q_3)e^{-\Phi} + {1\ov 3}
Q_4 e^{\Phi+2\l}
\ov {2 \a_1^2\a_2^2\a_3^2}} \ , 
\label{Killing} 
\\
\frac {1}{\a_3} \frac {d \a_3}{d \r} \! & = \! & \frac {g}{24} e^{- \Phi} ( 2 e^{2 \l} + e^{-4 \l} ) + 
\frac {1}{6g} e^{\Phi-2\l} \left( \frac {5}{\a_3^2} - \frac {1}{\a_1^2} - 
\frac {1}{\a_2^2} \right) 
\nonumber \\ 
& - & {(\a_1^2 Q_1 + \a_2^2 Q_2 - \a_3^2 Q_3)e^{-\Phi} + {1\ov 3}
Q_4 e^{\Phi+2\l}
\ov {2 \a_1^2\a_2^2\a_3^2}} \ , 
\nonumber \\
\frac {d \l}{d \rho} \! & = \! & \frac {g}{6} e^{-\Phi} ( e^{- 4 \l} - e^{2 \l} ) 
+ \frac {1}{3g} e^{\Phi - 2 \l} 
\left( \frac {1}{\a_1^2} + \frac {1}{\a_2^2} 
+ \frac {1}{\a_3^2} \right) + {Q_4 e^{\Phi+2\l} \ov {3 \a_1^2\a_2^2\a_3^2}} \ , 
\nonumber \\
\frac {d f}{d \r} \! & = \! & \frac {1}{3} \frac {d \Phi}{d \r} 
+ {2 (\a_1^2 Q_1 + \a_2^2 Q_2 + \a_3^2 Q_3)e^{- \Phi} + 2 Q_4 e^{\Phi+2\l}
\ov {3\a_1^2\a_2^2\a_3^2}}  
\nonumber\ .
\ea
Furthermore, we also obtain from the gravitino variation along $\r$ a 
differential equation yielding the $\r$-dependence of the 
Killing spinor as $\e=e^{f/2} \e_0\ ,$ where $\e_0$ 
is a constant spinor subject to the projections \eqn{projections}.
In fact, this functional form of the Killing spinor can be deduced just from 
the supersymmetry algebra.
  
We also note that the system (\ref{Killing}) also includes the case 
when the six-cycle is taken to be $S^2 \times \mathbb{CP}^2$ ;
this corresponds to setting equal radii for two of the spheres.
We only have to adjust the overall scale in the metric 
for $\mathbb{CP}^2$ to have scalar curvature equal to that of the metric
for the undeformed 
$S^2\times S^2$, which, in our normalization, equals four.
  
In order to solve the system \eqn{Killing} it is useful to redefine 
variables through  
\ba
&& dr  =  e^{-\Phi/3} d \r \ , \qq a_i = \a_i \, e^{-\Phi/3} \ ,\qq i=1,2,3\ ,
 \nonumber \\ 
&& a_4 = e^{\l + 2 \Phi/3} \ , \qq\ a = e^{- 2 \l + 2 \Phi/3} \ , 
\qq A = f-{\Phi\ov 3}\ . 
\label{ab}
\ea
In these variables the metric, when lifted to eleven dimensions, 
takes the form (in what follows we have conveniently 
set $g=2$; it can be reinstalled or taken to any value after appropriate 
rescalings)
\be
ds_{11}^2 = - e^{2 A}dt^2 + dr^2  + a_1^2 d\Om^2_2 +  a_2^2 d{\bar \Om}^2_2 + 
a_3^2 d{\tilde \Om}^2_2 + 
a_4^2 d{\hat \Om}^2_2 
+ a^2 \left( \hat{\s}_3 - \s_3 - \bar{\s}_3 - \tilde{\s}_3 \right)^2 \ ,
\label{lift}
\ee
where the $\hat{\s}_i$'s are left-invariant Maurer--Cartan $SU(2)$ one-forms 
satisfying as a triplet the conditions 
$d \hat{\s}_i = \frac {1}{2} \e_{ijk} \hat{\s}_j \wedge \hat{\s}_k$. The 
Killing spinor can also be lifted from eight to eleven dimensions 
through $\e_{11} = e^{-\Phi/6}\e=e^{A/2} \e_0$, while the eleven-dimensional four-form field 
strength corresponding to the uplift of \eqn{flux} is given by 
\ba
F_{0712} & = & {Q_1\ov  a_2^2 a_3^2 a_4^2 a} \ ,
\qq 
F_{07\bar{1}\bar{2}}  \ = \ {Q_2\ov  a_1^2 a_3^2 a_4^2 a}  \ ,
\nonumber\\
F_{07\tilde{1}\tilde{2}} & = & {Q_3\ov  a_1^2 a_2^2 a_4^2 a}\ ,\qq
F_{07\hat{1}\hat{2}} \ =  \ {Q_4\ov  a_1^2 a_2^2 a_3^2 a}\ . 
\label{lift4}
\ea
After the redefinitions \eqn{ab}, 
the system \eqn{Killing} becomes simpler,
\be
a_1 \frac {da_1}{dr}  =  \frac {a}{2} -{1\ov 3}{1\ov 
a_1^2 a_2^2 a_3^2 a_4^2 a} (-2 a_1^2 Q_1 + a_2^2 Q_2 + a_3^2 Q_3 + a_4^2 Q_4)  
\ , 
\label{ABCD1}
\ee
plus three more equations following from cyclic permutations in $1,2,3,4$,
as well as
\be
\frac {d a}{dr}  =  1- \frac {a^2}{2 } \left({1\ov a_1^2} 
+ {1\ov a_2^2} + {1\ov a_3^2} + {1\ov a_4^2} \right)- 
{1\ov 3}{1\ov a_1^2 a_2^2 a_3^2 a_4^2} (a_1^2 Q_1 + a_2^2 Q_2 + a_3^2 Q_3 + a_4^2 Q_4) \ , 
\nonumber
\label{ABCD}
\ee
whose solution determines the conformal factor as
\be
\frac {dA}{dr}  =   \frac {2}{3} \frac {1}{a_1^2 a_2^2 a_3^2 a_4^2 a} 
(a_1^2 Q_1 + a_2^2 Q_2 + a_3^2 Q_3 + a_4^2 Q_4) \ .
\label{coonn}
\ee
The above system should be supplemeted by the 
zero total charge condition \eqn{zzerr} and obvisously shares 
an $S_4$ permutation invariance originating from the equivalence of
all four two-spheres in the background \eqn{lift} and \eqn{lift4}.

We also note that having unequal radii for the spheres in the six-cycle provides the possibility to 
perform the limit where the radius of one of the two-spheres tends to infinity. This would 
amount to the growth of two flat coordinates; for instance 
for very large $a_4$ it is easily seen that 2+1 Lorentz invariance is restored and 
the system \eqn{ABCD1}-\eqn{coonn} (and the equivalent in \eqn{Killing}) 
becomes that in \cite{HS2}. In the absence of flux the metric (\ref{lift}) 
becomes $ds_{1,2}^2+ds_8^2$, where $ds_8^2$ is the metric of a Calabi--Yau fourfold, thus reducing 
from $SU(5)$ to $SU(4)$ holonomy. 
Similarly, one can go from $SU(4)$ to $SU(3)$ holonomy by blowing up 
one of the remaining two-spheres. 

Turning on fluxes as in \eqn{lift4} corresponds, from a string theory point
of view, to turning on D2-brane charges and forming 
a D2-D6 bound state with the entire spatial part of the world-volume
wrapped on the spheres. This interpretation is compatible with the fact that 
no additional projection is required as compared with the zero flux case and 
the amount of supersymmetry preserved is the same. 
We have been unable to solve the system \eqn{ABCD}-\eqn{coonn}, unlike the 
similar case in \cite{HS2}, in the presence of fluxes. In that respect note
that having non-vanishing fluxes is inconsistent with demanding simple 
solutions with equal radii, i.e. $a_1=\cdots =a_4$.

In subsection 3.1 we will show how to derive \eqn{ABCD1}-\eqn{coonn}
with \eqn{zzerr} directly using eleven-dimensional supergravity.

\subsection{$SU(5)$ holonomy}
  
Consider the case with vanishing fluxes, 
$Q_i=0$, with $i=1,2,3,4$. In this case the general solution to the 
system \eqn{ABCD1}-\eqn{ABCD} is (the conformal factor equals unity
in this case) 
\ba
 a_1^2 =  R^2 + l_1^2\ , \quad a_2^2 = R^2 + l_2^2 \ ,\quad 
a_3^2 = R^2 + l_3^2\ , \quad a_4^2 =  R^2, \quad a^2 = R^2 U^2(R) \ ,
\label{zero}
\ea
where
\be
U^2(R) = \frac {12 R^6 + 15 C_1 R^4 + 
20 C_2 R^2 + 30 C_3 + 12 C/R^4}
{30 (R^2 + l_1^2 ) ( R^2 + l_2^2 ) (R^2+l_3^2)} \ ,
\label{urr}
\ee
with the constants $C_i$, $i=1,2,3$ expressed as symmetric 
homogeneous polynomials, up to cubic order, in the $l_i^2$'s
\be
C_1= l_1^2 + l_2^2+l_3^2\ ,\qq
C_2 =l_1^2 l_2^2  + l_2^2 l_3^2  + l_3^2 l_1^2\ ,\qq 
C_3 = l_1^2 l_2^2 l_3^2\ .
\label{urr1}
\ee
The relation of the two variables $r$ and $R$ is via the differential
\be
dr={2\ov U(R)} dR\ .
\label{drdr}
\ee
Here we have denoted four of the 
constants of integration by $l_1,l_2,l_3 $ and $C$, 
and we have absorbed the fifth one by an appropriate shift in the variable 
$R$.\footnote{
The symmetry between all 
two-spheres can be manifestly restored in the solution \eqn{zero} if
we make the variable shift $R^2\to R^2+l_4^2$ and simultaneously 
redefine $l_i^2\to l_i^2-l_4^2$ for $i=1,2,3$.}
We can also see that in this case $e^{2\Phi}=R^3 U(R)$, $f=\Phi/3$ and
$A=0$. Hence the lifted eleven-dimensional solution in \eqn{lift} 
factorizes into the time coordinate and a Calabi--Yau fivefold with metric 
\be
ds_{10}^2 = dr^2  + a_1^2 d\Om^2_2 +  a_2^2 d{\bar \Om}^2_2 + 
a_3^2 d{\tilde \Om}^2_2 + a_4^2 d{\hat \Om}^2_2 
+ a^2 \left( \hat{\s}_3 - \s_3 - \bar{\s}_3 - \tilde{\s}_3 \right)^2 \ .
\ee
The asymptotic behaviour for large values of $R$ takes the universal form
\be
ds^2_{10} \, \simeq \, 10 dR^2 + R^2 ds_9^2\ ,\qq
 {\rm as}\quad R\to \infty\ 
\label{assy}
\ee
and it describes a cone whose base is given by
the nine-dimensional metric 
\be 
ds^2_9 =  d\Om_2^2 + d{\bar \Om}^2_2 + d{\tilde \Om}^2_2
+ d{\hat \Om}^2_2  + {2\ov 5} (\hat \s_3-\s_3-\bar \s_3-\tilde \s_3)^2\ .
\label{coonne}
\ee
This is an Einstein space obeying $R_{ij}={4\ov 5} g_{ij}$.
In fact it is the symmetric coset space 
\be
{SU(2)\times SU(2)\times SU(2)\times SU(2)\ov U(1)\times U(1)\times U(1)}\ ,
\ee
where the embedding of the $U(1)$'s into the Cartan subalgebra in the 
numerator is diagonal. We have been unable to find a previous reference and a 
nomenclature for it in the literature, but this is simply a higher dimensional analog 
of the largest similar case appearing as a Freund--Rubin compactification 
\cite{Freund} of eleven-dimensional supergravity \cite{11sugra} in four dimensions,
i.e. $Q^{1,1,1}$ \cite{FRE}. Extending that nomenclature we will refer to 
this space as $\Pi^{1,1,1,1}$.

In fact, \eqn{coonne}  is an exact solution for all values of $R$ as it can 
also be obtained by letting 
$l_1=l_2=l_3=C=0$ in the general solution. 
However, extending \eqn{coonne} to the 
interior is problematic because we reach a singularity at $R=0$, 
where the fiber and the $S^2$'s collapse to a point. 
Resolving the singularity to avoid this collapse requires that 
we turn on some of the different
moduli parameters which also determine the behavior of the solution in the
interior. In the following we further analyze for two different 
cases the solution for generic 
ranges of the parameters and show that indeed the 
singularity can be resolved in a manner similar
to that in \cite{HS2}.

\no
\underline{$l_1=l_2=l_3=0$}: In this case, when the constant $C\geq 0$,
the variable $R\geq 0$ and then the 
manifold is singular at $R=0$. If, however, $C=-\r_0^{10}<0$, where $\r_0$ 
is a 
real positive constant, then the variable $R\geq \r_0$ and 
the metric takes the simple form
\ba
ds_{10}^2 & = & {10 dR^2\ov 1-\r_0^{10}/R^{10}} + R^2 \left(d\Om^2_2
+d\bar \Om^2_2 + d\tilde \Om^2_2 + d\hat \Om^2_2\right)
\nonumber\\
&& +\ {2\ov 5} R^2
\left(1-\r_0^{10}/R^{10}\right)
(\hat \s_3-\bar \s_3-\tilde \s_3-\hat \s_3)^2\ .
\ea
Near $R=\r_0$ we change to a new radial variable $\tau=2\sqrt{\r_0(R-\r_0)}$ 
and find the behavior
\be
ds^2_{10} \simeq  a^2 ( d\Om_2^2 + d{\bar \Om}^2_2 + d{\tilde \Om}^2_2  
+ d{\hat \Om}^2_2)  \ 
  + \ d\tau^2 + \tau^2 (\hat \s_3-\s_3-\bar \s_3-\tilde \s_3)^2\ , 
\:\: {\rm as} \:\: \tau\to 0\ .
\label{bsoldd}
\ee
Therefore, near $\tau=0$ (or equivalently $R=\r_0$) and for
constant $\th$ and $\phi$, as well as for the corresponding barred, tilded
and 
hatted angles, the metric behaves as $dt^2 + t^2 d\hat \psi^2$
which shows that $t=0$ is a {\it bolt} singularity \cite{Gibbons}
which is removable provided that the periodicity 
of the angle $\hat{\psi}$ is restricted to $0\leq \hat \psi <2 \pi$.
Then the space becomes topologically 
$S^2\times S^2 \times S^2\times S^2\times  \mathbb{R}^2$ and the full
solution interpolates between this space for $R\to a$ and the ten-dimensional
space \eqn{assy} for $R\to \infty$. However, the latter is now a cone with
base $\Pi^{1,1,1,1}/\mathbb{Z}_2$ due to the above discrete identification.

\no
\underline{$l_1^2> 0,\ l_2^2>2\ {\rm and}\ l_3^2> 0$}: 
In this case, when the constant $C>0$, the 
variable $R\geq 0$ and there is a singularity at $R=0$.
If, however, $C=0$ then we have the behavior
\ba
ds^2_{10} & \simeq &   l_1^2  d\Om_2^2 + l_2^2 d{\bar \Om}^2_2 
+ l_3^2 d{\tilde \Om}^2_2 +
4 dR^2 + R^2 d{\hat \Om}^2_2 +
R^2 (\hat \s_3-\s_3-\bar \s_3-\tilde \s_3)^2\ ,
\nonumber\\
&& {\rm as}\quad R\to 0\ .
\label{bsolt}
\ea
Hence, for constant $\th,\phi$, $\bar\th,\bar\phi$ and 
$\tilde \th,\tilde\phi$ 
the metric 
behaves as $4 dR^2+ R^2(\hat\s_1^2 + \hat\s_2^2 + \hat\s_3^2) $ which shows
that we simply have a coordinate singularity in the polar coordinate system
on an $\IR^4$ centered at $R=0$. This is the so called {\it nut} singularity \cite{Gibbons}, 
which is removable by adding the point $R=0$ and changing to Cartesian 
coordinates. 
Therefore near $R=0$ the manifold becomes topologically 
$S^2\times S^2 \times S^2 \times \mathbb{R}^4$. Then the full
solution interpolates between this space for $R\to 0$ and the ten-dimensional
flat space in \eqn{assy} for $R\to \infty$.
If $C<0$ then there is an $R_0$ such that $U(R_0)^2=0$ 
(we take the largest root of this sixth order, in $R_0^2$, algebraic equation) 
and therefore 
we have that $R\geq R_0$. 
Changing to a new radial variable $\tau=2\sqrt{R_0(R-R_0)}$ we find the 
behavior
\ba
ds^2_{10} & \simeq & (R_0^2+l_1^2) d\Om_2^2 + (R_0^2+l_2^2) d{\bar \Om}^2_2
+ (R_0^2+l_3^2) d{\tilde \Om}^2_2
+ R_0^2 d{\hat \Om}^2_2 
\nonumber\\
&&   +  \, \, d\tau^2 + \tau^2 
(\hat \s_3-\s_3-\bar \s_3)^2\ ,\qq 
 {\rm as}\quad  \tau \to 0\ .
\label{jkoldd}
\ea
Hence the behavior is similar to that found before in \eqn{bsoldd}, with a 
removable {\it bolt} singularity at $\tau =0$.


\section{$SU(n\!+\!1)$ holonomy}
                           
The case of D6-branes wrapped on six-cycles with 
$S^2 \times \bar{S}^2 \times \tilde{S}^2$ 
topology we have considered in this paper is a generalization 
of that of D6-branes wrapped on $S^2 \times \bar{S}^2$ 
four-cycles, leading to $SU(4)$ holonomy \cite{GM,HS2}, or $S^2$ two-cycles corresponding 
to the resolved \cite{EN} or deformed conifold \cite{EPR3}. These 
constructions can be extended to an arbitrary even number of dimensions
$2n+2$ and $SU(n\!+\!1)$ holonomy. Of course for $n\geq 5$ we give up the
brane description of the underlying geometry. 
Consider the metric
\be
ds_{2n+2}^2 = dr^2 + \sum_{i=1}^n a_i^2 d \Omega_{2,i}^{2} + a^2 \Bigl(d\psi + \sum_{i=1}^n \sigma_{3,i}\Bigr)^2 \ ,
\label{sun}
\ee
containing a set of $n$ spheres, with $d\Om^2_{2,i}=\s_{1,i}^2+\s_{2,i}^2$ and
where the relation 
$d\s_{3,i}=\s_{1,i}\wedge \s_{2,i}$ defines the $\s_{3,i}$'s.

In the natural frame
\be 
e^{1}_i=a_i \s_{1,i}\ ,\qq e^{2}_i=a_i \s_{2,i}\ ,\qq
e^{2n+1}=a \Bigl(d\psi + \sum_{i=1}^n \sigma_{3,i}\Bigr) \ ,\qq e^{2n+2}=dr\ ,
\ee
the spin connection in a quite obvious notation reads
\ba
&& \om^{12}_{(i)}=\s_{3,i}-{a^2\ov 2 a_i^2} 
\Bigl(d\psi + \sum_{i=1}^n \sigma_{3,i}\Bigr) \ ,\:\:\:\:
\om^{2n+1\, 2n+2}={da\ov dr}
\Bigl(d\psi + \sum_{i=1}^n \sigma_{3,i}\Bigr) \ ,
\nonumber\\
&&\om^{1\, 2n+2}_{(i)} \! = {da_i\ov dr} \s_{1,i}\ ,
\:\: \om^{2\, 2n+2}_{(i)} \! = {da_i\ov dr} \s_{2,i}\ ,\:\: 
\om^{2n+1\, 1}_{(i)} \! ={a\ov 2 a_i} \s_{2,i}\ , \:\:
\om^{2 \, 2n+1}_{(i)} \! = {a\ov 2 a_i} \s_{1,i}\ .
\label{con32}
\ea

In order to generalize our construction to arbitrary even dimensions we 
will first extend the projections on the spinor as 
\be
\G^{1}_{(1)} \G^2_{(1)}\e =  \ldots = \G^{1}_{(n)} \G^{2}_{(n)} \e  
= -\G^{2n+1}\G^{2n+2}\e\ ,
\label{pprr}
\ee
representing, in total, $2^{n}$ independent conditions.
The metric \eqn{sun}
will admit a covariantly constant spinor provided that the
first order system of differential equations\footnote{This system was also 
studied in \cite{Cvetic}.}
\ba
a_i \frac {da_i}{dr} & = & \frac {a}{2} \ , \quad i=1,\ldots,n \ , 
\nonumber \\ 
\frac {d a}{dr}  & = & 1- \frac {a^2}{2 } \sum_{i=1}^n {1\ov a_i^2} \ ,
\label{sunsys}
\ea
is obeyed. The form of the covariantly constant spinor turns out to be
$\e=e^{\G^{12}_{(1)}\psi/2}\e_0$, with $\e_0$ a constant spinor subject to the
same projections as in \eqn{pprr}.
These projections leave two independent components for the spinor, 
so that we are left with an $SU(n+1)$ holonomy metric. 

The general solution to the system (\ref{sunsys}) is
\ba
dr & =& {2\ov U(R)} dR \ , \quad a^2 = R^2 U^2(R) \ , \nonumber \\
a_i^2 & = & R^2 + l_i^2\ , 
\quad \hbox{with } i=1,\ldots,n  \hbox{ and }l_n^2=0\ , 
\ea
where
\be
U^2(R) = \frac {2}{\prod_{i=1}^{n-1} (R^2+l_i^2)} 
\Big[ \frac {C/R^4}{n+1} + \sum_{i=0}^{n-1} \frac {R^{\, 2i}}{i+2} S_{n-i-1}(l^2) \Big] \ ,
\ee
with $C$ being a constant and $S_m(l^2)$ a symmetric homogeneous 
polynomial in the $l_i^2$'s, of degree $m$, with $m=0,1, \ldots, n-1$, 
\be
S_m(l^2) = 
\! \! \sum_{i_1<\ldots<i_m=1}^{n-1} l_{i_1}^2 l_{i_2}^2 \ldots l_{i_m}^2 \ .
\ee
The asymptotic behaviour of \eqn{sun} as $R \rightarrow \infty$ is that of
a cone with metric
\be
ds_{2n+2}^2 = 2(n+1) \, dR^2 + R^2 ds_{2n+1}^2 \ ,
\ee
and base the ($2n+1$)-dimensional Einstein space 
\be
ds_{2n+1}^2 = \sum_{i=1}^n d\Omega_{2,i}^2 + \frac {2}{n+1} 
\Bigl(d\psi + \sum_{i=1}^n \s_{3,i}\Bigr)^2 \ ,
\ee
with $R_{ij} = \frac {n}{n+1} g_{ij}$. This is the symmetric coset space
\be
\frac {\overbrace{SU(2) \times SU(2) \times \ldots 
\times SU(2)}^{n \hbox{ \footnotesize{factors}}}}
{\underbrace{U(1) \times \ldots 
\times U(1)}_{n-1 \hbox{ \footnotesize{factors}}}} \ ,
\ee
and we can denote it, by extending our previous notation corresponding to 
$n=4$, as $\Pi^{1,1,\ldots,1}$, with $n$ indices.
In the case where all $l_i=0$ and $C=-\r_0^{2n+2}<0$ we have
\be
U^2(R) = \frac {2}{n+1} \left( 1 - \frac {\r_0^{2n+2}}{R^{2n+2}} \right) \ .
\ee
Therefore near $R=\r_0$ we find a removable {\it bolt} singularity with 
$S^2_{(1)} \times S^2_{(2)} \times 
\ldots S^2_{(n)} \times \mathbb{R}^2$ topology. 
When $C=0$, and all $l_i^2 > 0$, a removable {\it nut} 
singularity arises near $R=0$, with $S^2_{(1)} \times S^2_{(2)} \times 
\ldots S^2_{(n-1)} \times \mathbb{R}^4$ topology. 
  
Obviously the cases with $n\geq 5$ do not have an 
interpretation in terms of branes and therefore seem to be of no direct 
interest to string and M-theory. Nevertheless, we find it quite interesting 
that the general structure is extendable for any $n$ and we believe that the
explicit forms of the metrics we have presented will be useful in general.

\subsection{Non-vanishing flux from eleven-dimensional supergravity}

It is worth examining the case of non-vanishing flux with background given
by \eqn{lift} and \eqn{lift4} directly using eleven-dimensional supergravity.
In particular, it is worth seeing how the condition 
\eqn{zzerr} arises in this approach.
The Killing spinor equation is
\be
\del_\m\e +{1\ov 4} \om_\m^{ab}\G_{ab}\e -{1\ov 288}\left(
F_{\n\r\l\s} \G^{\n\r\l\s}{}_\m -8 F_{\m\n\r\l} \G^{\n\r\l}\right)\e = 0\ .
\ee
The non-zero components of the spin connection are given, after 
appropriate relabeling, by \eqn{con32},
where the general case should be specialized to $n=4$ and supplemented with
$\om^{0,10}=e^A dA/dr\, dt$. Similarly, the appropriate projections are
given by \eqn{pprr} (with $n=4$).
It turns out that the Killing spinor is $\e=e^{A/2} e^{\ha \G_{12}\psi}\e_0$,
provided that the system of equations \eqn{ABCD1}-\eqn{coonn} is obeyed (with
charges rescaled by the factor $-2$), with
no restriction, such as \eqn{zzerr}, for the charges. 
However, we have to make sure that the equation of motion for the three-form 
gauge potential is obeyed. For the ansatz \eqn{lift4} the contribution
to it from the Chern--Simons term in the action vanishes. Hence, we have that
\be
{1\ov \sqrt{-g}}\del_\m \left(\sqrt{-g} F^{\m\n\r\l}\right)= 0 \ .
\ee
For every choice of $\n,\r$ and $\l$ this is trivially satisfied with the 
background ansatz \eqn{lift} and \eqn{lift4}. However, satisfying the equation
for $(\n,\r,\l)=(t,r,\psi)$ requires imposing the zero total charge
condition \eqn{zzerr}.


\bs\bs

\centerline {\bf Acknowledgments}

R.H. acknowledges the financial support provided through the European
Community's Human Potential Programme under contract HPRN-CT-2000-00131
``Quantum Structure of Space-time'', the Swiss Office for Education 
and Science and the Swiss National Science Foundation. 
  
K.S. acknowledges the financial support provided through the European
Community's Human Potential Programme under contracts HPRN-CT-2000-00131
``Quantum Structure of Space-time'' and 
HPRN-CT-2000-00122 ``Superstring Theory'', by the Greek State
Scholarships Foundation under the contract IKYDA-2001/22 ``Quantum Fields
and Strings'', as well as NATO support by a Collaborative Linkage Grant under the
contract PST.CLG.978785 ``Algebraic and Geometrical Aspects of Conformal
Field Theories and Superstrings''. He also acknowledges the hospitality and
financial support of the TH-Division of CERN, 
where a substantial part of this work was done.


\end{document}